\documentclass[aps, pra, reprint, superscriptaddress, nofootinbib]{revtex4-1}
\pdfoutput=1

\usepackage[final]{microtype}
\usepackage[USenglish]{babel}
\usepackage[utf8x]{inputenc}
\usepackage[T1]{fontenc}
\usepackage[version=4]{mhchem}
\usepackage[colorlinks=true, linkcolor=blue, citecolor=blue, urlcolor=blue, bookmarks=false]{hyperref}
\usepackage{graphicx, csquotes, physics, siunitx, newtxtext, newtxmath}

\newcommand{\ie}{\textit{i.e.} }
\newcommand{\eg}{\textit{e.g.} }
\newcommand{\etal}{\textit{et al.}}

\newcommand{\ts}{\textsuperscript}

\newlength{\lineheight}
\DeclareMathOperator{\sinc}{sinc}

\begin{document}

    \title{Optimal Optomechanical Coupling Strength in Multi-Membrane Systems}

    \author{David C. Newsom}
        \affiliation{Department of Physics, University of California, Santa Barbara, Santa Barbara, California 93106, USA}
    \author{Fernando Luna}
        \affiliation{Department of Physics, University of California, Santa Barbara, Santa Barbara, California 93106, USA}
    \author{Vitaly Fedoseev}
        \affiliation{Huygens-Kamerlingh Onnes Laboratorium, Universiteit Leiden, 2333 CA Leiden, The Netherlands}
    \author{Wolfgang Löffler}
        \affiliation{Huygens-Kamerlingh Onnes Laboratorium, Universiteit Leiden, 2333 CA Leiden, The Netherlands}
    \author{Dirk Bouwmeester}
        \affiliation{Department of Physics, University of California, Santa Barbara, Santa Barbara, California 93106, USA}
        \affiliation{Huygens-Kamerlingh Onnes Laboratorium, Universiteit Leiden, 2333 CA Leiden, The Netherlands}

    \begin{abstract}
        We calculate the optomechanical coupling strength for a multi-membrane in a cavity system. The optimal configuration for an array of $N$ membranes placed near the center of a cavity is identified. This results in a coupling strength much greater than previously proposed multi-membrane configurations. We find that the coupling strength scales exponentially with the number of membranes until saturating due to complete localization of the field within the array. Furthermore we explore two sources of loss, those due to light absorption within the membrane(s) and leakage through the cavity's end-mirrors, and evaluate how they affect the possibility of achieving strong coupling.
    \end{abstract}

    \maketitle

\section{Introduction}
\label{Introduction:Sec}

    Optomechanical systems span a wide range of configurations and sizes, giving access to different parameters and operational regimes. The membrane-in-the-middle (MiM) configuration, first introduced by Thompson \etal~\cite{thompsonStrongDispersiveCoupling2008}, has proved to be versatile in supporting the integration of various technologies such as phononic~\cite{tsaturyanUltracoherentNanomechanicalResonators2017} and photonic~\cite{norteMechanicalResonatorsQuantum2016} crystals to enhance the mechanical quality factor and the reflectivity of the membrane, stress engineering~\cite{norteMechanicalResonatorsQuantum2016} to further increase mechanical $Q$ and reduce mechanical noise, and metal coating~\cite{yuanSiliconNitrideMembrane2015, andrewsBidirectionalEfficientConversion2014} to couple to microwave resonators.

    For a single membrane configuration, Jayich \etal~\cite{jayichDispersiveOptomechanicsMembrane2008} demonstrated an approach for calculating the first order optomechanical coupling strength and the characteristics of the optical cavity as a function of the membrane's position and reflectivity. First and second-order optomechanical coupling rates were then derived for a two-membrane system~\cite{bhattacharyaMultipleMembraneCavity2008} in order to couple the electromagnetic field to collective modes of the mechanics and achieve optomechanically mediated long-range interactions between different mechanical elements. Subsequent work has invested the possibility of enhanced optomechanical coupling in a two-membrane system~\cite{piergentiliTwomembraneCavityOptomechanics2018, liCavityModeFrequencies2016,gartnerIntegratedOptomechanicalArrays2018} and in a system of $N$ identical, evenly spaced membranes~\cite{xuerebStrongCouplingLongRange2012, xuerebCollectivelyEnhancedOptomechanical2013}.

    Each of the above mentioned works have followed the same general method for calculating the coupling strength. A resonance condition for the cavity frequencies is first derived and then linearized about the desired configuration. However the complexity of the resonance condition increases significantly as the number of membranes grows, making this method ill-suited for designing many-membrane configurations with large coupling strengths. Here we present an alternative method that provides direct insight into the multi-membrane coupling strength.

    Sec.~\ref{SingleMembrane:Sec} of this paper describes this alternative method for determining the linear coupling strength in a single membrane cavity. As is intuitively clear, one finds that the coupling strength of a membrane to an optical mode depends solely on the difference in intensity of the standing waves on either side of the membrane. Analyzing the dependance of these region-specific intensities on the large (\ie cavity-length) and small (\ie wavelength) scale position of the membrane we show how to maximize the membrane's coupling strength.

    In Sec.~\ref{MultiMembranes:Sec}, we extend our method to calculate the coupling strengths of multiple, identical membranes within a cavity. We discuss the coupling strength of the fields to the collective mechanical modes and introduce the collective coupling strength of the entire system. Using insights gained by our analysis, we construct a many-membrane configuration whose collective coupling strength scale exponentially with the number of membranes before saturating. The saturation occurs due to the nearly complete localization of the field within the array of membranes. We confirm our analytical results by numerical simulations, solving for the resonance condition for the same configuration.

    In Sec.~\ref{DecayRate:Sec}, we calculate the loss rate due to leakage through the mirrors and absorbtion in the membrane(s). We show that through the use of multiple membranes it is possible to eliminate leakage through the mirrors, however absorption necessarily scales with the linear coupling of the system, resulting in fundamental limits for the enhancement of the coupling relative to the decay rate of the cavity for a given material.

\section{Single Membrane}
\label{SingleMembrane:Sec}

     We begin by treating the cavity as one dimensional and the mirrors as perfectly reflecting. We orient the axis of our cavity along the $z$-axis with the origin positioned in the cavity's center. The membranes are initially treated as slabs of lossless dielectric material. Their positions and material properties are described by the cavity's dielectric function $\epsilon(z)$. This, together with boundary conditions at the end-mirrors, determines the cavity's resonant modes $\phi(z)$ and their associated frequencies $\omega$ through the eigenvalue equation~\cite{cheungNonadiabaticOptomechanicalHamiltonian2011}
    \begin{equation}\label{SingleMembrane:Eq:EigenvalueEquation}
        \frac{1}{\epsilon(z)}\pdv[2]{z}\phi(z)=-\frac{\omega^2}{c^2}\phi(z).
    \end{equation}

    \begin{figure}
        \centering
        \includegraphics{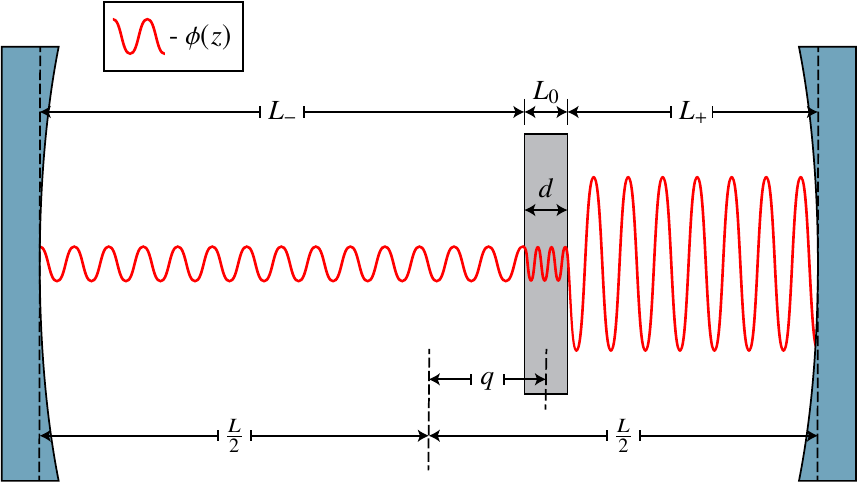}
        \caption{Schematic of a single membrane system.}\label{SingleMembrane:Fig:SingleMembrane_Schematic}
    \end{figure}

    Initially, we restrict our analysis to a system of a single membrane interacting with a single optical mode. The dielectric function for such a system is given by
    \begin{equation}\label{SingleMembrane:Eq:DielectricFunction}
        \epsilon(z) = \begin{cases}
                            n^2, & \; \abs{z-q}<\frac{d}{2} \\[0.5\lineheight]
                            1, &\; \abs{z-q}>\frac{d}{2}
                      \end{cases}
    \end{equation}
    where $n$, $d$, and $q$, are the membrane's (real) index of refraction, thickness, and center-of-mass (CoM) coordinate, respectively.

    The membrane divides the cavity into three distinct dielectric regions as shown in Fig.~\ref{SingleMembrane:Fig:SingleMembrane_Schematic}. Within each of the regions, the solution to Eq.~\ref{SingleMembrane:Eq:EigenvalueEquation} is a standing wave. The optical mode of our system may then be written in the piecewise form
    \begin{equation}\label{SingleMembrane:Eq:ModeFunction}
        \phi(z) = \sqrt{\frac{\hbar \omega}{\epsilon_0 V_\mathrm{Cav}}A}\cdot   \begin{cases}
                                                                                    \sqrt{I_-}\cos(\frac{\omega}{c}\qty(z-\frac{L}{2})+\theta_-) \\[\lineheight]
                                                                                    \sqrt{\frac{I_0}{n^2}}\cos(\frac{n\omega}{c}\qty(z-\qty(q-\frac{d}{2}))+\theta_0) \\[\lineheight]
                                                                                    \sqrt{I_+}\cos(\frac{\omega}{c}\qty(z-\qty(q+\frac{d}{2}))+\theta_+),
                                                                                \end{cases}
    \end{equation}
    where $I_-$, $I_0$, and $I_+$ are the dimensionless \emph{regional intensities} associated with the standing waves to the left, inside, and to the right of the membrane. We enforce that the regional intensities sum to unity and let the overall normalization of the mode be determined by the dimensionless \emph{absolute intensity} $A$.

    The regional intensities and the regional phases $\theta_i$ are related through the matching conditions at the material interfaces inside the cavity. We make this relationship more explicit in Sec.~\ref{SingleMembrane:SubSubSec:RegionalIntensities} when discuss the dependency of the membrane's optomechanical coupling strength on its position and show that the regional and absolute intensities exert the dominant influence on the optomechanical interaction.

\subsection{Single-Photon Coupling Strength}
\label{SingleMembrane:SubSec:SinglePhotonCouplingStrength}

    The interaction between the mechanical motion of the membrane and the optical mode in our cavity is characterized to first order by the \emph{single-photon coupling strength} $g_0$. It is defined here as
    \begin{equation}\label{SingleMembrane:Eq:CouplingStrengthDefinition}
        g_0 \equiv q_\mathrm{zpf}\pdv{\omega}{q},
    \end{equation}
    where $q_\mathrm{zpf}$ is the zero-point fluctuation in the membrane's position and $\pdv{\omega}{q}$ is the shift in the mode's resonant frequency due to a displacement of the membrane\footnote{Note that our sign convention differs from that found in \cite{aspelmeyerCavityOptomechanics2014}.}.

    The quantity $\pdv{\omega}{q}$ may be calculated directly from the averaged radiation pressure felt by the membrane (see Sup. Info. Sec.~\ref{SupplmentaryInfo:SubSec:RadiationPressureForce}) and is proportional to the difference in the standing wave intensities on either side of the membrane, resulting in:
    \begin{equation}\label{SingleMembrane:Eq:CouplingStrength_SimpleForm}
      g_0 = q_\mathrm{zpf}\cdot A\frac{\omega}{L}\qty(I_+-I_-)
    \end{equation}

    The advantage of writing the coupling strength in the form of Eq.~\ref{SingleMembrane:Eq:CouplingStrength_SimpleForm} over other forms such as those in \cite{johnsonPerturbationTheoryMaxwell2002} is the absence of any dependency on the field values at the membrane. In Eq.~\ref{SingleMembrane:Eq:CouplingStrength_SimpleForm}, the only dependency is on the regional intensities, $I_+$ and $I_-$, and the absolute intensity $A$. We now focus on analyzing how the membrane's position controls these quantities. Once this is known, it is straightforward to determine the membrane position which yields the maximum value for the coupling strength.

\subsection{Determining the Field Intensities}
\label{SingleMembrane:SubSec:FieldIntensties}

\subsubsection{Positioning on the Wavelength Scale: Regional Intensities}
\label{SingleMembrane:SubSubSec:RegionalIntensities}

    In order to determine the relationship between $I_-$, $I_0$, and $I_+$ in Eq.~\ref{SingleMembrane:Eq:ModeFunction}, we decompose the waves in each region into their plane wave components and use the transfer matrix formalism.

    Letting $\varphi_\pm$ denote the phase of the right/left standing waves at each surface of the membrane, the regional intensities $I_+$ and $I_-$ satisfy
    \begin{equation}\label{SingleMembrane:Eq:TransferMatrix}
        \sqrt{I_+}  \begin{pmatrix}
                        e^{i\varphi_+} \\
                        e^{-i\varphi_+}
                    \end{pmatrix} = \frac{i}{\sqrt{1-r^2}}  \begin{pmatrix}
                                                                e^{i\theta_r}   &   -r              \\
                                                                r               &   -e^{-i\theta_r}  \\
                                                            \end{pmatrix} \cdot \sqrt{I_-}  \begin{pmatrix}
                                                                                                e^{i\varphi_-} \\
                                                                                                e^{-i\varphi_-}
                                                                                            \end{pmatrix},
    \end{equation}
    where $r$ and $\theta_r$ are the magnitude and phase of the membrane's amplitude reflectivity.

    We may solve for the membrane's \emph{intensity ratio}, $\gamma$:
    \begin{equation}\label{SingleMembrane:Eq:IntensityRatio}
        \gamma \equiv \frac{I_+}{I_-} = \frac{1-2r\cos(2\varphi_-+\theta_r)+r^2}{1-r^2}
    \end{equation}

    We see that $\gamma$ is periodic under half-wavelength translations of the membrane from its sinusoidal dependance on the field's phase. This periodicity allows us to tune the regional intensities by making sub-wavelength adjustments to the membrane's position, irrespective of its location in the cavity. The intensity ratio across the membrane is therefore a wavelength scale effect of the membrane's position.

    Using the same method, we may find a relationship similar to Eq.~\ref{SingleMembrane:Eq:IntensityRatio} relating $I_0$ to $I_-$ and $I_+$.

\subsubsection{Positioning on the Cavity-Length Scale: Absolute Intensity}
\label{SingleMembrane:SubSubSec:AbsoluteIntensity}

    The absolute intensity of the mode is determined by the fact that the field energy associated with a photon is $\hbar\omega$. This normalization condition fixes $A$, giving
    \begin{equation}\label{SingleMembrane:Eq:AbsoluteIntensity_Normalization}
        A = \frac{1}{I_-\frac{L_-}{L}+I_0\frac{L_0}{L}+I_+\frac{L+}{L}}.
    \end{equation}
    We note that $L_0=d$ is independent of $q$, leaving the regional intensities and the remaining length factors $\frac{L_\pm}{L}$ as the terms still dependant on the membrane's position. While the regional intensities are sensitive to wavelength scale changes in the membrane's position, the length factors are insensitive to such changes for cavities in which $L\gg \lambda$. This difference in scale-relative sensitivity allows us to treat these two quantities as independent. Once the regional intensities are set, the absolute intensity is entirely a cavity-length scale effect of the membrane's position.

    It is common for the membrane's thickness to be much smaller than both the cavity's length and the optical wavelength, in which case we may take the \emph{thin approximation} and neglect the field energy within the membrane when normalizing the mode. This is equivalent to setting $\frac{L_0}{L}=0$ in Eq.~\ref{SingleMembrane:Eq:AbsoluteIntensity_Normalization}.

\subsection{Maximizing the Coupling}
\label{SingleMembrane:SubSec:MaximizingCoupling}

    We are now ready to determine the optimal position for a single membrane within a cavity in order to maximize the coupling strength. At the wavelength scale, we want to position the membrane to maximize the difference in regional intensities across the membrane, $\abs{I_+-I_-}$. As noted in Sec.~\ref{SingleMembrane:SubSubSec:RegionalIntensities} this coupling strength maximum occurs when the membrane's intensity ratio achieves one of its extremal values $\frac{1 \pm r}{1 \mp r}$. In the following we assume thin membranes and take $\gamma = \frac{1+r}{1-r}$ so that $I_+>I_-$.

    Our choice of $\gamma$ fixes the regional intensities within the cavity, allowing us to solve for the absolute intensity $A$ for a general membrane position $q$. The resulting optomechanical coupling strength is
    \begin{equation}\label{SingleMembrane:Eq:CouplingStrength_GeneralPosition}
        g_0(q) = q_\mathrm{zpf}\frac{2\omega r}{L}\cdot\frac{1}{1-2r\frac{q}{L}} \equiv g_1\cdot\frac{1}{1-2r\frac{q}{L}}.
    \end{equation}
    We emphasize that this $q$ is approximate up to sub-wavelength adjustments. In actuality, fixing the regional intensities limits shifting the membrane's position to half-integer multiples of the wavelength. However, these increments are small enough relative to the length of cavity that we may treat $\frac{q}{L}$ as a continuous parameter in Eq.~\ref{SingleMembrane:Eq:CouplingStrength_GeneralPosition}.

    The quantity $g_1 \equiv q_\mathrm{zpf}\frac{2\omega r}{L}$ represents the maximal value for a single membrane's coupling strength when positioned near the center of the cavity (\ie $\frac{q}{L} \approx 0$). We use $g_1$ as a reference to determine the factor by which the coupling strength is enhanced due to the membrane's position. By shortening the cavity length $L$, we restrict the field to a smaller region and both $g_1$ and $g_0$ increase as a result.

    When the membrane is placed near the left mirror, the region of high intensity occupies the majority of the cavity. Such a configuration requires a lower absolute intensity to satisfy the normalization condition, and the coupling strength is \emph{reduced} relative to that at the center of the cavity. Conversely, when the membrane is placed near the right mirror, the region of high intensity is restricted to a minority of the cavity and the coupling strength is \emph{enhanced} relative to that at the center of the cavity. We conclude that the coupling strength can generally be increased by reducing the length of the high intensity regions within the cavity.

\section{Extension to Multiple Membranes}
\label{MultiMembranes:Sec}

    We now examine systems of multiple membranes. For simplicity, we treat the case of identical membranes. For a system of $N$ membranes, the cavity's dielectric function is shown in Eq.~\ref{MultiMembranes:Eq:DielectricFunction} where the $q_i$ terms are the CoM coordinates of the different membranes with $i$ from 1 to $N$ (left to right).

    \begin{equation}\label{MultiMembranes:Eq:DielectricFunction}
        \epsilon(z) =   \begin{cases}
                            n^2, & \abs{z-q_1}<\frac{d}{2},\, \abs{z-q_2}<\frac{d}{2},\,\dots \\[0.5\lineheight]
                            1, & z<q_1-\frac{d}{2}, q_1+\frac{d}{2}<z<q_2-\frac{d}{2},\,\dots
                        \end{cases}
    \end{equation}

    The cavity is now partitioned into $2N+1$ regions and the mode function $\phi(z)$ remains in the form of a standing wave within each region. We use the same convention as that of Eq.~\ref{SingleMembrane:Eq:ModeFunction} and enforce the mode's regional intensities to sum to unity with the normalization being set by the absolute intensity.

\subsection{Individual Couplings}
\label{MultiMembranes:SubSec:IndividualCouplings}

    The cavity mode's resonant frequency $\omega$ now depends on the $N$ different CoM coordinates $q_i$, resulting in each membrane having its own individual coupling strength
    \begin{equation}\label{MultiMembranes:Eq:IndividualCouplingStrenghDefinintion}
        g^{(i)} \equiv q_\mathrm{zpf}\pdv{\omega}{q_i}.
    \end{equation}
    This is analogous to Eq.~\ref{SingleMembrane:Eq:CouplingStrengthDefinition} for the single-membrane case, so we may write
    \begin{equation}\label{MultiMembranes:Eq:IndividualCouplingStrength_SimpleForm}
        g^{(i)} = q_\mathrm{zpf}\cdot A\frac{\omega}{L}\qty(I_+^{(i)}-I_-^{(i)}),
    \end{equation}
    where $I_\pm^{(i)}$ is the regional intensity to the left/right of the $i$\ts{th} membrane\footnote{This notation for the regional intensities, while mirroring the convention used for the single membrane system, is overcomplete as $I_+^{(i)}$ and $I_-^{(i+1)}$ both refer to the same region within the cavity.}.

    The regional intensities are related by each membrane's intensity ratio
    \begin{equation}\label{MultiMembranes:Eq:IntensityRatio}
        \gamma_i\equiv\frac{I_+^{(i)}}{I_-^{(i)}} = \frac{1-2r\cos(2\varphi_-^{(i)}+\theta_r)+r^2}{1-r^2}.
    \end{equation}

    Just as in the single membrane case, the regional intensities are determined by the position of the membranes at the wavelength scale. The mode function $\phi(z)$ must still be normalized such that the photon energy is $\hbar\omega$, hence the absolute intensity is given by
    \begin{equation}\label{MultiMembranes:Eq:AbsoluteIntensity_Normalization}
        A = \qty[\sum_{i=1}^{N}I_i\frac{L_i}{L}]^{-1}.
    \end{equation}
    With the regional intensities fixed, the absolute intensity is again set by the cavity-length scale position of the membranes.

    Because $g^{(i)}$ is a generalized form of that for a single membrane system, we follow the same approach for maximizing the individual coupling strengths. For each membrane, we position it at the wavelength scale such that its intensity ratio $\gamma_i$ is extremized. Once the regional intensities are set, we adjust the configuration at the cavity-length scale so that the high intensity regions occupy a small portion of the cavity.

    The presence of multiple membranes offers greater control over the intensity profile within the cavity. Using $I_+^{(i)}=\gamma_i I_-^{(i)}$ and $I_-^{(i)}=I_+^{(i-1)}$, we express the individual coupling strengths as
    \begin{equation}\label{MultiMembranes:Eq:IndividualCoupling_NeighborForm}
        g^{(i)} = q_\mathrm{zpf}\cdot AI_+^{(i-1)}\cdot\frac{\omega}{L}\qty(\gamma_i-1).
    \end{equation}
    By controlling the intensity ratio $\gamma_{i-1}$ of the membrane's left neighbor, we may increase $I_+^{(i-1)}$ independently of $\gamma_i$. This creates a region of high intensity and effectively increases the absolute intensity felt by membranes in this region. By compounding this effect across several membranes, very high intensities are possible and membranes placed within these regions can have individual coupling strengths significantly exceeding what would be possible for a single membrane system.

\subsection{Collective Coupling}
\label{MultiMembranes:SubSec:CollectiveCoupling}

    One motivation to study an array of multiple membranes is to investigate the possible enhancement of the optomechanical properties through collective motion of the membranes~\cite{bhattacharyaMultipleMembraneCavity2008, piergentiliTwomembraneCavityOptomechanics2018, liCavityModeFrequencies2016, xuerebStrongCouplingLongRange2012, xuerebCollectivelyEnhancedOptomechanical2013}. A multi-membrane system can exhibit numerous styles of collective motion. We refer to a specific style of collective motion as a collective mechanical mode. For example, a two membrane system possesses a center-of-mass mode, where the membranes are synchronized and move in the same direction, and a breathing mode, where the membranes are anti-synchronized and move in opposite directions. To describe an arbitrary collective mode, we introduce a mode coordinate $u$ and specify how the individual position of each membrane evolves with $u$. In general, a membrane's position $q_i$ is a linear function of the mode coordinate,
    \begin{equation}\label{MultiMembranes:Eq:CollectiveModeProfile}
        q_i = a_i u + b_i.
    \end{equation}
    The set of weights $a_i$ determines the relative motion of each membrane (\ie the mode's mechanical profile) and the constants $b_i$ determine their resting positions. We emphasize that the set of $a_i$ does not encode physical information about the system; it specifies the collective mode of the array under consideration.

    The coupling strength for a collective mode is defined analogously to that of the a multi-membrane system as
    \begin{equation}\label{MultiMembranes:Eq:ModeCollectiveCoupling_Definition}
        g^{(u)}\equiv u_\mathrm{zpf}\pdv{\omega}{u} = u_\mathrm{zpf}\sum_{i=1}^{N}a_i\pdv{\omega}{q_i}.
    \end{equation}
    We normalize the mode by requiring $\sum a_i^2=1$, rendering the mode's zero point fluctuations equivalent to that of a single membrane.

    The collective mode $u_c$ which has the largest coupling strength to the field has a mechanical profile determined by the individual coupling strengths of the system, $a_i \propto g^\qty(i)$. The coupling of this mode defines the system's \emph{collective coupling strength}
    \begin{equation}\label{MultiMembranes:Eq:SystemCollectiveCoupling_Definition}
        g_c\equiv\sqrt{\sum_{i=1}^{N}\qty(g^{(i)})^2}.
    \end{equation}
    The system's collective coupling strength is the highest coupling strength achievable within a given configuration.

    In the case of identical membranes, the system's first order optomechanical Hamiltonian reduces to a single term that couples $u_c$ to the optical mode~\cite{xuerebCollectivelyEnhancedOptomechanical2013}. Any dynamical effects resulting from this interaction (\eg backaction) will be the same as those of a system consisting of a single mechanical element coupled to an optical mode with strength $g_c$. Such effects are beyond the scope of this article, so we refer the reader to \cite{aspelmeyerCavityOptomechanics2014} for more information.

\subsection{Maximizing the Collective Coupling}
\label{MultiMembranes:SubSec:Maximizing the Collective Coupling}

    It is straightforward to construct a system of membranes whose collective coupling far exceeds that of a single membrane. As discussed in Sec.~\ref{MultiMembranes:SubSec:IndividualCouplings}, a strongly coupled configuration is one in which the intensity gradient across each membrane is maximized while the field energy is confined to a small region of the cavity. A simple example of such a configuration is an evenly spaced $N$-membrane array, where each element's intensity ratio saturates to $\Gamma\equiv\frac{1+r}{1-r}$. The regional intensity then grows by a factor of $\Gamma$ across each element, making the total intensity ratio of the array $\Gamma^N$. This configuration requires the array to be positioned near one of the end mirrors to confine the field energy as discussed in Sec.~\ref{SingleMembrane:SubSec:MaximizingCoupling}. An explicit treatment of this configuration is given in Supplementary~Information~Sec.~\ref{SupplmentaryInfo:SubSec:MultiMembranesMirror}.

\begin{figure}
        \centering
        \includegraphics{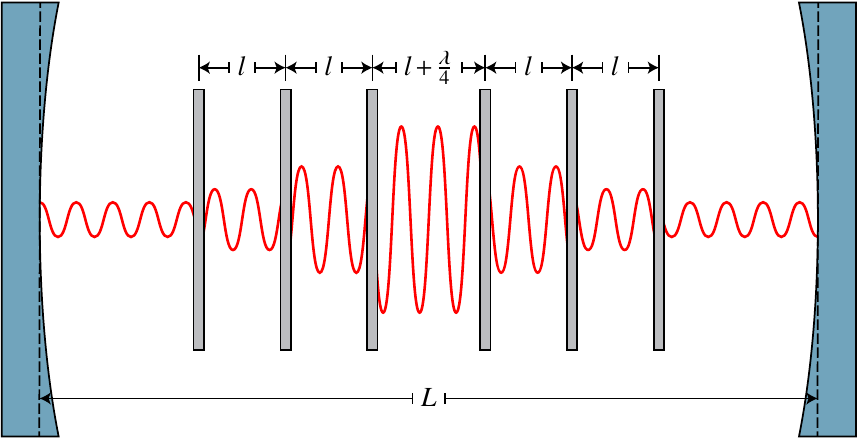}
        \caption{Optimal configuration for an array of membranes centered in the cavity.}\label{MultiMembranes:Fig:MultiMembrane_Schematic}
    \end{figure}

    Alternatively, we construct an array with a similar structure, but which localizes the field in the center of the cavity. Letting $N$ be even, the membranes are placed symmetrically about the center of the cavity, spaced by a distance $l$, subject to sub-wavelength adjustments. The field intensity increases from either side by a factor of $\Gamma$, until reaching the center of the array. This configuration is depicted for a system of six membranes in Fig.~\ref{MultiMembranes:Fig:MultiMembrane_Schematic}. For this system, the individual coupling strengths are given by
    \begin{equation}\label{MultiMembranes:Eq:CenterConfig_IndividualCouplings}
        \begin{split}
           g^{(i)} = \frac{1}{2}g_1 \cdot   &   \frac{\Gamma-1}{r+\qty(\Gamma^{\frac{N}{2}}-\qty(rN+1))\frac{l}{L}} \\[\lineheight]
                                    \times  &   \begin{cases}
                                                    \Gamma^{i-1},   &   1\leq i\leq\frac{N}{2} \\[0.5\lineheight]
                                                    -\Gamma^{N-i},  &   \frac{N}{2}< i\leq N.
                                                \end{cases}
        \end{split}
    \end{equation}
    The collective coupling strength is
    \begin{equation}\label{MultiMembranes:Eq:CenterConfig_CollectiveCoupling}
        g_c = g_1\sqrt{\frac{r}{2}}\cdot\frac{\sqrt{\Gamma^N-1}}{r+\qty(\Gamma^{\frac{N}{2}}-\qty(rN+1))\frac{l}{L}}.
    \end{equation}

\begin{figure}
        \centering
        \includegraphics{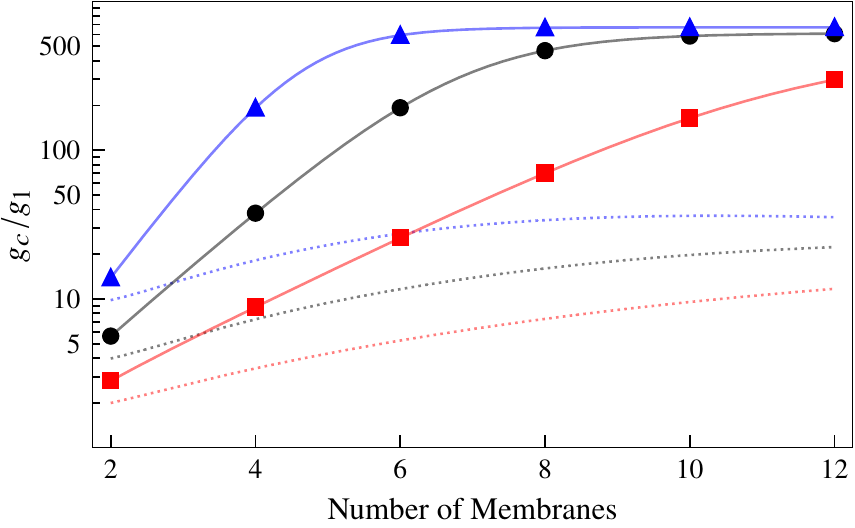}
        \caption{Collective coupling strength of the system depicted in Fig.~\ref{MultiMembranes:Fig:MultiMembrane_Schematic} with $\frac{L}{l}\approx \num{e3}$. The red (lower), black (middle), and blue (upper) plots correspond to a membrane reflectivity of \numlist{.5;.75;.9} respectively. The solid curves represent the theoretical values calculated from Eq.~\ref{MultiMembranes:Eq:CenterConfig_CollectiveCoupling} while the plotted points represent the coupling strength numerically calculated by solving for the resonance frequencies of the system as a function of the mode coordinate. All elements are treated as lossless. For comparison, the couplings of systems, with the same approximate spacing and membrane reflectivity, in the configuration proposed by \cite{xuerebCollectivelyEnhancedOptomechanical2013} are shown in dashed lines.}\label{MultiMembranes:Fig:CollectiveCoupling_Plot}
    \end{figure}

    As $N$ increases, so does the field energy trapped inside the array, increasing the coupling strength. In the limit of many membranes, the coupling strength saturates to
    \begin{equation}\label{MultiMembranes:Eq:CenterConfig_SaturatedCoupling}
        g_\mathrm{sat} = g_1 \cdot \sqrt{\frac{r}{2}}\frac{L}{l} = \sqrt{2r^3}\frac{q_\mathrm{zpf}}{l}\omega
    \end{equation}
    once the field energy is entirely localized within the array. Since there is no energy outside the array, this saturation value is independent of the length of the cavity. This is in contrast to the case of a single membrane, whose coupling depend on the cavity length. Fig.~\ref{MultiMembranes:Fig:CollectiveCoupling_Plot} shows that a configuration with small $\frac{l}{L}$ can result in a collective coupling strength that is orders of magnitude greater than that of a single membrane system.

    Both the exponential scaling with observed in Eq.~\ref{MultiMembranes:Eq:CenterConfig_CollectiveCoupling} and the saturated coupling strength $g_\mathrm{sat}$ exceed previously proposed multi-membrane configurations~\cite{xuerebCollectivelyEnhancedOptomechanical2013}. This is the result of choosing the membrane configuration which maximizes the individual coupling strengths in the system as well as localizing the majority of the field energy within the array. Furthermore, once the coupling is saturated, increasing the number of elements does not affect the coupling strength.

    The precise spacing of the membranes is found by determining the phase of the field required to set the regional intensities. The allowed values are
    \begin{equation}\label{MultiMembranes:Eq:CenterConfig_MembraneSpacing}
        l = \frac{\lambda}{2}\qty(\frac{3}{2}-\frac{\theta_r}{\pi}+n)\quad n\in\mathbb{N}.
    \end{equation}
    The spacing between the two innermost membranes must be extended by an addition quarter wavelength for the intensity profile to be symmetric within the array.

    For mirrors with an amplitude reflectivity of \num{1}, the resonant lengths of the cavity are
    \begin{equation}\label{MultiMembranes:Eq:CenterConfig_CavityLength}
        L = \qty(N-1)l+\lambda\qty(\frac{7}{4}-\frac{\theta_r}{2\pi}+n)\quad n\in\mathbb{N}.
    \end{equation}

    \begin{figure}
        \centering
        \includegraphics{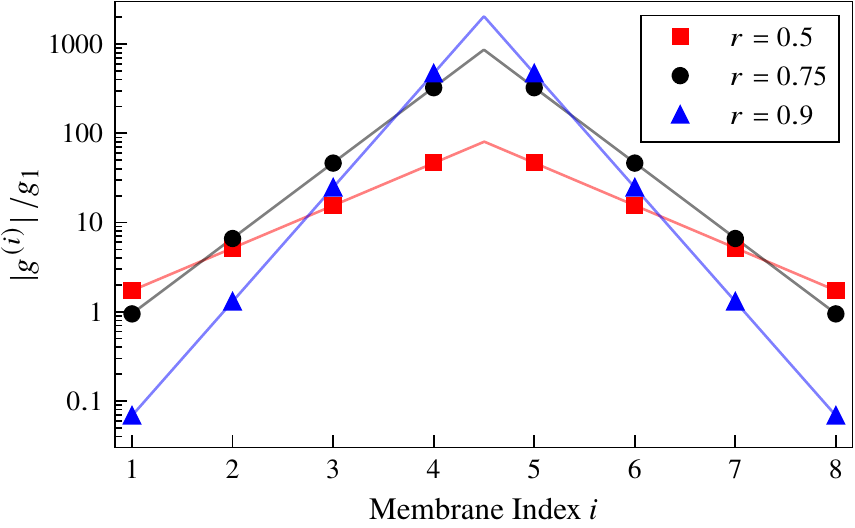}
        \caption{Individual coupling strengths of eight membranes in the configuration of Fig.~\ref{MultiMembranes:Fig:MultiMembrane_Schematic} with $\frac{L}{l}\approx \num{e3}$. These are directly proportional to the optical force acting on each membrane (see Eq.~\ref{SingleMembrane:Eq:CouplingStrength_SimpleForm}). The solid curves represent the theoretical values calculated from Eq.~\ref{MultiMembranes:Eq:CenterConfig_IndividualCouplings} while the plotted points represent the coupling strength numerically calculated by solving for the resonance frequencies of the system as a function of the mode coordinate. All elements are treated as lossless. }\label{MultiMembranes:Fig:IndividualCouplings_Plot}
    \end{figure}

\subsection{Extension to Non-Identical Membranes}
\label{MultiMembranes:SubSec:Non-IdenticalMembranes}

    Our method easily generalizes to handle the case of non-identical membranes. The only modification needed is in Eq.~\ref{MultiMembranes:Eq:IntensityRatio} when defining a membrane's intensity ratio. As each membrane now has potentially distinct material properties, $r$ and $\theta_r$ must now have membrane specific indices. One complication which may arise when analyzing systems with non-identical membranes is that the collective mechanical mode which directly couples to the field is not guaranteed to be a normal mode mechanical mode of the system. If this is the case, it will be mechanically coupled to other vibrational modes of the membrane array.

\section{Photon Decay Rate}
\label{DecayRate:Sec}

    Up to this point in our discussion, we have been assuming the membranes and end-mirrors to be lossless in order to simplify the associated calculations. We now include photon loss. The dissipation of light within the cavity is characterized by the \emph{photon decay rate} $\kappa$, defined as
    \begin{equation}\label{DecayRate:Eq:DecayRate_Definition}
        \kappa \equiv \frac{1}{\hbar\omega\qty(n+\frac{1}{2})}\ev**{\hat{W}}{n}.
    \end{equation}

    The operator $\hat{W}$ in Eq.~\ref{DecayRate:Eq:DecayRate_Definition} represents the power dissipated within the cavity. There are two sources of such dissipation which we consider. Correspondingly, $\hat{W}$, and by extension $\kappa$, may be decomposed into two terms as in Eq.~\ref{DecayRate:Eq:LossOperator_SourceDecomp}, where $\hat{W}_\mathcal{T}$ represents the energy lost through the end-mirrors and $\hat{W}_\sigma$ represents the energy lost through absorption within the membranes.

    \begin{equation}\label{DecayRate:Eq:LossOperator_SourceDecomp}
        \hat{W} = \hat{W}_\mathcal{T}+\hat{W}_\sigma
    \end{equation}

    We consider only the limiting case of weak losses, where any effect on the optical mode profile $\phi(z)$ is negligible. This then allows us to use the lossless multi-membrane extension of the mode profile (Eq.~\ref{SingleMembrane:Eq:ModeFunction}) in place of the more complex, albeit exact, mode profile for lossy systems in subsequent decay rate calculations. We will see that such an assumption holds quite well for realistic values of end-mirror transmission and membrane absorption when compared against numerical calculations.

\subsection{Transmissive Mirrors}
\label{DecayRate:SubSec:TransmissiveMirrors}

    In the context of our earlier single membrane system depicted in Fig.~\ref{SingleMembrane:Fig:SingleMembrane_Schematic}, now consider the case where the end-mirrors each possess a finite transmission coefficient $\mathcal{T} \ll 1$. The power which escapes the cavity through the end-mirrors is directly proportional to the power impinging on the mirrors from within the cavity (see Sup. Info. Sec.~\ref{SupplmentaryInfo:SubSec:TransmissiveMirrors}). The resulting contribution to the decay rate from the mirrors is
    \begin{equation}\label{DecayRate:Eq:MirrorDecayRate_General}
        \kappa_\mathcal{T} = \mathcal{T}\cdot\frac{c}{2L}\cdot A\qty(I_-+I_+).
    \end{equation}

    Eq.~\ref{DecayRate:Eq:MirrorDecayRate_General} remains valid for arbitrary multi-membrane configurations, where $I_\pm$ now represents the regional intensities of the field adjacent to the end-mirrors. In the configuration detailed in Sec.~\ref{MultiMembranes:SubSec:Maximizing the Collective Coupling}, we have
    \begin{equation}\label{DecayRate:Eq:MirrorDecayRate_CenterConfig}
        \kappa_\mathcal{T} = \frac{c}{L}\cdot\frac{r\mathcal{T}}{r+\qty(\Gamma^{\frac{N}{2}}-\qty(rN+1))\frac{l}{L}}.
    \end{equation}
    It can be seen that the mirrors' contributions to the photon decay rate diminish as the number of membranes is increased. This is to be expected because higher membrane numbers reduce the field intensity near the mirrors.

\subsection{Absorptive Membranes}
\label{DecayRate:SubSec:AbsorptiveMembranes}

    As in the previous section, we use the single membrane case depicted in Fig.~\ref{SingleMembrane:Fig:SingleMembrane_Schematic} before generalizing our result to multiple membranes. Absorptive membranes may be characterized by introducing an imaginary component $\tilde{n}$ to the refractive index of the material forming the membrane~\cite{jayichDispersiveOptomechanicsMembrane2008}. In order to calculate the power dissipation within such a material, it is simplest to treat $\tilde{n}$ as arising from a finite conductivity $\sigma$. The loss operator associated with absorption is found by integrating the Ohmic heating induced by this conductivity through the volume of the membrane (see Sup. Info. Sec.~\ref{SupplmentaryInfo:SubSec:AbsorptiveMembranes}). The membrane's contribution to the decay rate is
    \begin{align}
        \kappa_\sigma &= 4\frac{\tilde{n}c}{n^2 L}\cdot AI_0\cdot \xi(\varphi, \theta_0). \label{DecayRate:Eq:MembraneDecayRate_General} \\[0.5\lineheight]
        \xi(\varphi, \theta) &= \frac{1}{2}\qty(\varphi+\sin(\varphi)\cos(2\theta+\varphi)) \label{DecayRate:Eq:MembraneDecayRate_GeometricFactor}
    \end{align}
    The quantity $\xi$ in Eq.~\ref{DecayRate:Eq:MembraneDecayRate_GeometricFactor} characterizes the field volume within the membrane, where $\varphi\equiv\frac{n\omega d}{c}$ is a phase that depends on the thickness of the membrane. For a system of multiple membranes, each membrane contributes a term similar to Eq.~\ref{DecayRate:Eq:MembraneDecayRate_General} to the total decay rate of the cavity.

\begin{figure}
      \centering
      \includegraphics{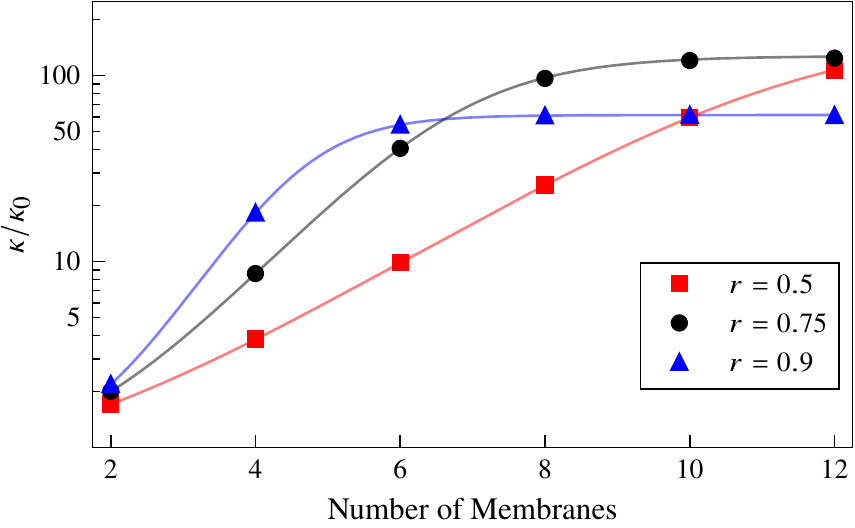}
      \caption{Photon decay rate of the system depicted in Fig.~\ref{MultiMembranes:Fig:MultiMembrane_Schematic} with transmissive mirrors and absorptive membranes. The membranes possess an extinction coefficient $\tilde{n}=\num{e-5}$ and the mirrors have a transmission coefficient of $\mathcal{T}=\num{5e-5}$. The solid curves represent the theoretical values predicted by Eqs.~\ref{DecayRate:Eq:MirrorDecayRate_CenterConfig}~and~\ref{DecayRate:Eq:MembraneDecayRate_CenterConfig} while the plotted points represent the value for the cavity's linewidth obtained from numerically calculating the full-width-half-maximum of the system's transmission peak. The dimensions of the cavity are the same as those used for Fig.~\ref{MultiMembranes:Fig:CollectiveCoupling_Plot}. As a reference scale, we use the decay rate of an empty cavity, $\kappa_0 \equiv \mathcal{T}\cdot\frac{c}{L}$.}\label{DecayRate:Fig:AbsorptiveMembranesDecayRate_Plot}
    \end{figure}

    In the configuration of Sec.~\ref{MultiMembranes:SubSec:Maximizing the Collective Coupling}, the decay rate due to absorption within the membranes is
    \begin{equation}\label{DecayRate:Eq:MembraneDecayRate_CenterConfig}
        \kappa_\sigma = \tilde{n}\cdot\frac{c}{L}\cdot \chi \cdot \frac{\Gamma^\frac{N}{2}-1}{r+\qty(\Gamma^{\frac{N}{2}}-\qty(rN+1))\frac{l}{L}}
    \end{equation}
    \begin{equation}\label{DecayRate:Eq:MembraneDecayRate_CenterConfig_ChiFactor}
        \chi \equiv \varphi\qty[\qty(1+n^{-2})+r\qty(1-n^{-2})\cos(\theta_r)]+\frac{2r}{n}\sin(\theta_r).
    \end{equation}
    Results from numerical calculations for the absorptive membranes in this configuration are shown in Fig.~\ref{DecayRate:Fig:AbsorptiveMembranesDecayRate_Plot}.

    In contrast to the mirrors' contribution to the decay rate of Eq.~\ref{DecayRate:Eq:MirrorDecayRate_CenterConfig} which decreases to zero as the number of membranes is increased, the membranes' contribution, shown in Eq.~\ref{DecayRate:Eq:MembraneDecayRate_CenterConfig}, increases in the same manner as the collective coupling of the system in Eq.~\ref{MultiMembranes:Eq:CenterConfig_CollectiveCoupling}. As we show in the next section, the contribution of absorption to the photon decay rate necessarily scales with the coupling strength.

\subsection{Limitations of Absorptive Membranes}
\label{DecayRate:SubSec:LimitationsAbsorptiveMembranes}

    An attractive feature of multi-membrane configurations is the possibility that the enhanced coupling strength achieved will allow systems to reach the so called \enquote{single-photon strong coupling regime}~\cite{piergentiliTwomembraneCavityOptomechanics2018, liCavityModeFrequencies2016, xuerebStrongCouplingLongRange2012, xuerebCollectivelyEnhancedOptomechanical2013}, which occurs when the coupling strength exceeds the photon decay rate. In this section we derive the relationship between the coupling strength of a membrane and the decay rate caused by the membrane's light absorption and reveal a fundamental limitation systems with absorptive membranes have in achieving the single-photon strong coupling regime.

\begin{figure}
        \centering
        \includegraphics{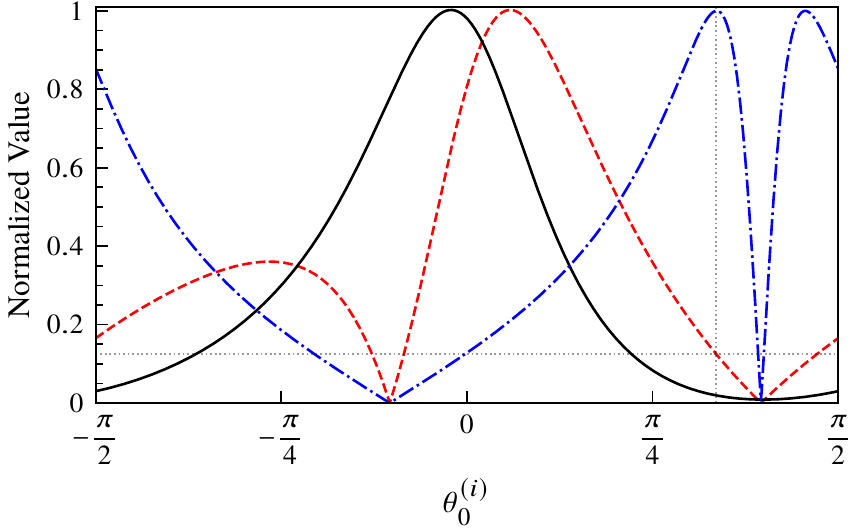}
        \caption{Coupling $\abs*{g^{(i)}}$ (dashed red), absorption decay rate $\kappa_\sigma^{(i)}$ (solid black), and coupling efficiency $\eta^{(i)}$ (dash-dotted blue) as a function of the field phase for \SI{50}{\nano\meter} thick \ce{Si3N4} membrane and laser wavelength of \SI{1064}{\nano\meter}. These plots are normalized with respect to their maximum value. The dotted grey lines show that at peak efficiency, the membrane experiences approximately \SI{12.5}{\percent} of its maximum coupling. We treat the intensity $I_-^{(i)}$ external to the membrane as constant, so that the intensity $I_0^{(i)}$ internal to the membrane has implicit dependance on the field's phase.}\label{DecayRate:Fig:CouplingEfficiency_Plot}
    \end{figure}

    We define a membrane's coupling efficiency $\eta^{(i)}$ as the ratio of the membrane's coupling strength to its decay rate contribution from absorption:
    \begin{equation}\label{DecayRate:Eq:CouplingEfficiency_General}
        \begin{split}
           \eta^{(i)} &\equiv \frac{\abs{g^\qty(i)}}{\kappa_\sigma^\qty(i)} \\[0.5\lineheight] &=\frac{q_\mathrm{zpf}}{\lambda}\cdot\frac{\pi(n^2-1)}{\tilde{n}}\cdot\frac{\abs{\sinc(\varphi)\sin(2\theta_0^{(i)}+\varphi)}}{1+\sinc(\varphi)\cos(2\theta_0^{(i)}+\varphi)}
        \end{split}
    \end{equation}
    With the material and geometry of the membrane fixed, the efficiency of a membrane is independent of the field intensity within the membrane and depends soley on the field's phase $\theta_0^{(i)}$, which may be set by choice. The behavior of the coupling strength, decay rate, and efficiency as a function of the field's phase is shown in Fig.~\ref{DecayRate:Fig:CouplingEfficiency_Plot} for a \ce{Si3N4} membrane. It can be seen that maximizing the coupling efficiency comes with a significant reduction in the overall magnitude of the coupling.

    In a system of a single membrane where the only source of loss is through absorption, strong single-photon coupling is achieved when the membrane possesses a coupling efficiency exceeding unity. Optimizing over the field's phase, we find that the highest coupling efficiency achievable by a membrane of fixed material and thickness is
    \begin{equation}\label{DecayRate:Eq:CouplingEfficiency_Max}
        \eta_\mathrm{max} = \frac{q_\mathrm{zpf}}{\lambda}\cdot\frac{\pi(n^2-1)}{\tilde{n}}\cdot\frac{\abs{\sinc(\varphi)}}{\sqrt{1-\sinc^2(\varphi)}}.
    \end{equation}
    Requiring Eq.~\ref{DecayRate:Eq:CouplingEfficiency_Max} to exceed unity is a necessary condition for the possibility of achieving strong coupling which depends \emph{only} on the material properties and geometry of the membrane and \emph{not} on its position within the cavity.

    For multiple membranes, Eq.~\ref{DecayRate:Eq:CouplingEfficiency_Max} still holds as an upper bound on the ratio of the collective coupling strength and the total photon decay rate of a system of $N$ identical membranes\footnote{In the non-identical case, an upper bound on this ratio is given by the highest individual coupling efficiency by one of the types of membranes present.} (see Sup. Info. Sec.~\ref{SupplmentaryInfo:SubSec:CollectiveEfficiencyBound}). We note than that in systems with very weakly absorbing membranes the mirrors contribution to the decay rate $\kappa_\mathcal{T}$ may decrease faster than $\kappa_\sigma$ increases. In such a case, the use of multiple membranes would result in an overall decrease in the decay rate.

    We estimated the magnitude of a \SI{50}{\nano\meter} thick \ce{Si3N4} membrane's zero point motion to be on the order of \SI{e-15}{\meter}. With a nominal value of $\tilde{n}_{\ce{Si3N4}} \approx \num{e-5}$~\cite{jayichDispersiveOptomechanicsMembrane2008, sankeyStrongTunableNonlinear2010}, the maximum coupling efficiency of such a membrane with a laser of wavelength \SI{1064}{\nano\meter} is $\eta_\mathrm{max}^{\qty(\ce{Si3N4})}\approx\num{3e-3}$. This can be increased by thinning the membrane, but increasing the efficiency by three orders of magnitude would result in an impracticably thin membrane. This result demonstrates that creating a system of \enquote{slab} \ce{Si3N4} membranes which has strong single-photon coupling is not possible.

    Any method for achieving strong single-photon coupling, irrespective of configuration or number of membranes, will require a design approach where the individual membrane's maximum coupling efficiency exceed unity. Such design changes could be implemented by either constructing membranes of materials other than \ce{Si3N4} and/or constructing membranes with additional structure, such as referenced in \cite{norteMechanicalResonatorsQuantum2016, buiHighreflectivityHighQMicromechanical2012, stambaughMembraneinthemiddleMirrorinthemiddleHighreflectivity2015}. In the latter case, which modifies the profile of the field within the membrane, the expressions for absorbtion we derived in Sec.~\ref{DecayRate:SubSec:AbsorptiveMembranes} for simple \enquote{slab} membranes will require modification according to the specific structure imposed (see Sup. Info. Sec.~\ref{SupplmentaryInfo:SubSec:BeyondSimpleMembranes} for a discussion of how to treat such cases). The results of the previous sections which made no reference to the internal structure of the membrane remain valid with the proviso that the modified structure of the membrane still allows it to be characterized by a reflectivity $r$.

\subsection{Cooperativity}
\label{DecayRate:SubSec:Cooperativity}

    While systems with absorptive membranes are fundamentally limited in enhancing the ratio of the coupling and the photon decay rate, they may still provide a significant enhancement to the systems single-photon cooperativity
    \begin{equation}\label{DecayRate:Eq:Cooperativity_Definition}
        \mathcal{C}_0 = \frac{4g_c^2}{\kappa \Gamma_m},
    \end{equation}
    where $\Gamma_m$ is the mechanical damping rate of the membranes. For the case of identical membranes, the mechanical damping rate is independent of the mechanical mode chosen~\cite{xuerebStrongCouplingLongRange2012}. The cooperativity characterizes the efficiency with which cavity photons and phonons may be exchanged~\cite{aspelmeyerCavityOptomechanics2014} and is relevant to many optomechanical processes such as effective laser cooling~\cite{chanLaserCoolingNanomechanical2011, yuanLargeCooperativityMicrokelvin2015}.

    The enhancement of $\mathcal{C}_0$ in the configuration of Sec.~\ref{MultiMembranes:SubSec:Maximizing the Collective Coupling}, compared to that of a single, non-absorbing membrane in the center of a cavity, is
    \begin{equation}\label{DecayRate:Eq:CooperativityEnhancement}
        \frac{\mathcal{C}_0}{\mathcal{C}_1} = \frac{1}{2}\cdot\frac{r\mathcal{T}}{r\mathcal{T}+\tilde{n}\chi\qty(\Gamma^\frac{N}{2}-1)}\cdot\frac{\Gamma^N-1}{r+\qty(\Gamma^\frac{N}{2}-(rN+1))\frac{l}{L}}.
    \end{equation}
    \begin{equation}\label{DecayRate:Eq:SingleMemCooperativity}
        \mathcal{C}_1 \equiv \frac{4g_1^2}{\Gamma_m\kappa_0}
    \end{equation}
    In the limit of many membranes, Eq.~\ref{DecayRate:Eq:CooperativityEnhancement} saturates to
    \begin{equation}\label{DecayRate:Eq:CooperativitySaturation}
        \frac{\mathcal{C}_\mathrm{sat}}{\mathcal{C}_0} = \frac{r}{2}\cdot\frac{\mathcal{T}}{\tilde{n}\chi}\cdot\frac{L}{l}.
    \end{equation}
    A system of \SI{50}{\nano\meter} thick \ce{Si3N4} membranes can produce an enhancement of up to $\frac{\mathcal{C}_0}{\mathcal{C}_1}\approx\num{e4}$ when $\frac{L}{l} \approx \num{5e4}$ and $\mathcal{T}\approx \num{5e-5}$.

    This enhancement may be even higher for non-simple membranes whose absorption and reflectivity are superior to that of simple membranes. The equivalent expression to Eq.~\ref{DecayRate:Eq:CooperativitySaturation} for such membranes is
    \begin{equation}
        \frac{\mathcal{C}_\mathrm{sat}}{\mathcal{C}_0} = \frac{r}{2}\cdot\frac{\mathcal{T}}{\mathcal{A}}\cdot\frac{L}{l},
    \end{equation}
    where $\mathcal{A}$ is the membranes absorption coefficient. For membranes which possess a reflection and absorption coefficient of $r^2 = \SI{99.4}{\percent}$ and $\mathcal{A} = \num{e-7}$ respectively, the enhancement of the cooperativity in a system where $\mathcal{T}\approx \num{5e-5}$ and $\frac{L}{l}\approx \num{6.3e3}$ is $\frac{\mathcal{C}_0}{\mathcal{C}_1} \approx \num{1.3e6}$. This is an order of magnitude higher than previously proposed configurations with similar parameters~\cite{xuerebStrongCouplingLongRange2012}.

     \begin{figure}
      \centering
      \includegraphics{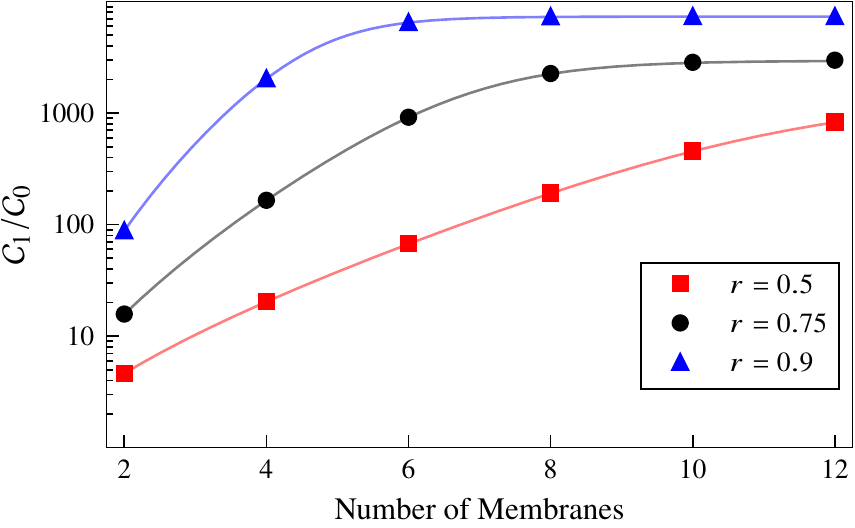}
      \caption{Cooperativity of the system depicted in Fig.~\ref{MultiMembranes:Fig:MultiMembrane_Schematic} with transmissive mirrors and absorptive membranes. The membranes possess an extinction coefficient $\tilde{n}=\num{e-5}$ and the mirrors have a transmission coefficient of $\mathcal{T}=\num{5e-5}$. The solid curves represent the theoretical values predicted by Eq.~\ref{DecayRate:Eq:CooperativityEnhancement} while the plotted points represent the results of numerical calculations. The dimensions of the cavity are the same as those used for Fig.~\ref{MultiMembranes:Fig:CollectiveCoupling_Plot}.}\label{DecayRate:Fig:Cooperativity_Plot}
    \end{figure}

\section{Conclusions}
\label{Conclusions:Sec}

    We have presented a derivation of the optomechanical coupling strength for a membrane-in-the-middle system and have identified that the coupling strength ultimately depends on the field intensity across the membrane. We have shown that this difference in field strength is determined by the position of the membrane at both the large and small scales, and that these two distinct effects can be controlled independently. Extending this analysis to a system with multiple membranes, we observed that a membrane in the array can experience an enhanced coupling strength resulting from the increased confinement of the field due to the other membranes. Our description of the collective mechanical modes included a heuristic presentation of the cavity configuration (in terms of membrane position) that yields the highest possible coupling strength. Our method showed that the collective coupling strength increases as more membranes are added until saturation is reached due to the complete localization of the field within the membrane array. Finally, we explored the effect of loss on the decay rate of the cavity and provided a necessary condition on membrane design for strong single-photon coupling. We found that for simple \ce{Si3N4} membranes with practical parameters this condition cannot be met. On the other hand, a considerable enhancement of the cooperativity is attainable.

\section{Supplementary Information}
\label{SupplementaryInfo:Sec}

\subsection{Calculating $g_0$ via Radiation Pressure Force}
\label{SupplmentaryInfo:SubSec:RadiationPressureForce}

    We compute $\pdv{\omega}{q}$ for the system of Fig.~\ref{SingleMembrane:Fig:SingleMembrane_Schematic}. Upon quantization, the mode frequency $\omega$ is directly proportional the energy levels of the field. We may employ the \emph{Hellman-Feynman theorem}~\cite{feynmanForcesMolecules1939} to calculate the first order correction to the these energies due a shift in the membrane's position $q$:
    \begin{equation}\label{SupplmentaryInfo:Eq:HellmanFeynmanTheorem}
        \pdv{\omega}{q} = \frac{\ev**{\pdv{\hat{\mathcal{H}}}{q}}{n}}{\hbar(n+\frac{1}{2})}
    \end{equation}

    \begin{equation}\label{SupplmentaryInfo:Eq:HamiltonianRadiationPressure}
        \hat{F}_z = -\pdv{\hat{\mathcal{H}}}{q}
    \end{equation}

    The radiation pressure force felt by the membrane is related to the field's Hamiltonian by Eq.~\ref{SupplmentaryInfo:Eq:HamiltonianRadiationPressure} and therefore, Eq.~\ref{SupplmentaryInfo:Eq:HellmanFeynmanTheorem} naturally leads to the standard interpretation of $g_0$ as a measure of the radiation pressure force per photon felt by the membrane as described in \cite{aspelmeyerCavityOptomechanics2014}. $\hat{F}_z$ may be calculated by enforcing momentum conservation for the membrane-field system via
    \begin{equation}\label{SupplmentaryInfo:Eq:RadiationForceStressTensor}
        \hat{F}_z = -\int_{V}\dd[3]{r}\frac{\epsilon(z)}{c^2}\pdv{\hat{S}_z}{t}+\int_{\partial V}\dd{A_i}\hat{\mathcal{T}}_{iz}.
    \end{equation}

    Here the integral is taken over \emph{any} volume $V$ enclosing the membrane. The quantity which appears in the first of Eq.~\ref{SupplmentaryInfo:Eq:RadiationForceStressTensor} is Poynting vector $\hat{S}_z$ and represents the field momentum in the region of integration. It is noteworthy that due to the \emph{Abraham-Minkowski controversy}~\cite{zangwillModernElectrodynamics2013} there is disagreement about the inclusion of $\epsilon(z)$ in the momentum density of the field. This debate is immaterial to our discussion, however, because the Poynting vector vanishes when averaged over a stationary state.

    It is the second term in Eq.~\ref{SupplmentaryInfo:Eq:RadiationForceStressTensor} that contributes significantly to $\hat{F}_z$. The divergence theorem has been used to convert this term to a surface integral of the Maxwell stress tensor $\hat{\mathcal{T}}$ and is interpreted as the momentum flux flowing through the boundary of $V$. It is readily evaluated by taking $\partial V$ over the outer surface of the membrane, so that the surface normal points parallel to the cavity axis and perpendicular to the electric and magnetic fields. The resulting force, averaged over a number state, is
    \begin{equation}\label{SupplmentaryInfo:Eq:AverageRadiationForce}
        \ev**{\hat{F}_z}{n} = \qty(n+\frac{1}{2})\cdot A\frac{\hbar\omega}{L}\qty(I_--I_+).
    \end{equation}

    This expression for $\ev*{\hat{F}_z}$ is quite similar to that for an end-mirror of an empty cavity~\cite{aspelmeyerCavityOptomechanics2014}; the only difference being the intensity factor $A\qty(I_--I_+)$. the origin of this factor can be understood qualitatively by observing that, in contrast to an end-mirror, the membrane has radiation impinging on both sides.

    Having calculated the radiation pressure force $\hat{F}_z$, $g_0$ assumes the simple form
    \begin{equation}\label{SupplementaryInfo:Eq:CouplingStrength_SimpleForm}
        g_0 = q_\mathrm{zpf}\cdot A\frac{\omega}{L}\qty(I_+-I_-).
    \end{equation}

\subsection{Multiple Membranes near a Mirror}
\label{SupplmentaryInfo:SubSec:MultiMembranesMirror}

\begin{figure}
          \centering
          \includegraphics{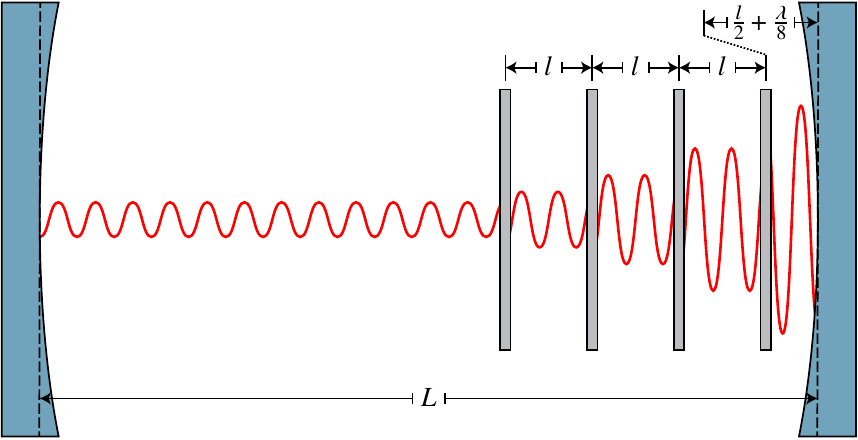}
          \caption{Optimal configuration for an array of membranes at the edge the cavity.}\label{SupplmentaryInfo:Fig:MultiMembrane_EndMirror_Schematic}
    \end{figure}

   Fig.~\ref{SupplmentaryInfo:Fig:MultiMembrane_EndMirror_Schematic} shows the configuration which results in maximal coupling when four membranes are positioned near an end-mirror of the cavity. The membranes are positioned such that the field intensity grows by a factor of $\Gamma\equiv\frac{1+r}{1-r}$ as one moves through the array. Unlike the configuration of Fig.~\ref{MultiMembranes:Fig:MultiMembrane_Schematic}, which had the field focused in the center of the array, the field is now focused outside the array in the region by the nearest end-mirror.

    Generalizing from four membranes to $N$ membranes, the individual and collective coupling strengths in this configuration are
    \begin{align}
      g^\qty(i) &= \frac{1}{2} g_1\cdot\frac{\Gamma-1}{r+\frac{1}{2}\qty(\Gamma^N-(2rN+1))\frac{l}{L}}\cdot\Gamma^{i-1} \label{SupplmentaryInfo:Eq:Mirror_IndividualCouplings}\\[\lineheight]
      g_c &= g_1\cdot\frac{\sqrt{r}}{2}\cdot\frac{\sqrt{\Gamma^{2N}-1}}{r+\frac{1}{2}\qty(\Gamma^N-(2rN+1))\frac{l}{L}}.\label{SupplmentaryInfo:Eq:Mirror_CollectiveCoupling}
    \end{align}
    In the limit of many membranes, the coupling saturates to
    \begin{equation}\label{SupplmentaryInfo:Eq:Mirror_SaturatedCoupling}
        g_\mathrm{sat} = g_1\cdot\sqrt{r}\frac{L}{l} = 2\sqrt{r^3}\frac{q_\mathrm{zpf}}{l}\omega.
    \end{equation}
    These results are very similar to those obtained in Sec.~\ref{MultiMembranes:SubSec:Maximizing the Collective Coupling}. The two notable differences are that the collective coupling grows as $\Gamma^N$ (rather than $\Gamma^\frac{N}{2}$ in Eq.~\ref{MultiMembranes:Eq:CenterConfig_CollectiveCoupling}) and that the limiting value $g_\mathrm{sat}$ is greater than Eq.~\ref{MultiMembranes:Eq:CenterConfig_SaturatedCoupling} by a factor of $\sqrt{2}$. These enhancements come at the cost of increased optical loss. In particular, the optical loss through the mirrors no longer vanishes in the limit of many membranes as it did for the configuration of Fig.~\ref{MultiMembranes:Fig:MultiMembrane_Schematic}.

    The precise membrane spacing $l$ is determined to be
    \begin{equation}\label{SupplmentaryInfo:Eq:Mirror_MembraneSpacing}
        l = \lambda\qty(\frac{3}{4}-\frac{\theta_r}{2\pi}+n)\quad n\in\mathbb{N}.
    \end{equation}
    For mirrors with an amplitude reflectivity of $1$, the resonant lengths of the cavity are
    \begin{equation}\label{SupplmentaryInfo:Eq:Mirror_ResonantCavityLength}
        L = \qty(N-\frac{1}{2})l+\frac{\lambda}{2}\qty(\frac{7}{4}-\frac{\theta_r}{2\pi}+n)\quad n\in\mathbb{N}.
    \end{equation}

\subsection{Loss from Transmissive Mirrors}
\label{SupplmentaryInfo:SubSec:TransmissiveMirrors}

    The power which escapes the cavity through the end-mirrors is directly proportional to the power impinging on the mirrors from within the cavity. We compute the impinging power from the plane waves in the region adjacent to the mirror. For simplicity, we consider the left end-mirror in the configuration of Fig.~\ref{SingleMembrane:Fig:SingleMembrane_Schematic}. The field operators for the forward and backward prorogating plane waves are
    \begin{align}
        \hat{E}_-^{(\pm)} &= -\frac{i}{2}\sqrt{AI_-\frac{\hbar\omega}{\epsilon V_\mathrm{Cav}}}\qty[\hat{a}^\dagger e^{\mp i\qty(\frac{\omega}{c}z+\theta_-)}-\hat{a}e^{\pm i\qty(\frac{\omega}{c}z+\theta_-)}] \label{SupplmentaryInfo:Eq:ElectricPlaneWave}\\[\lineheight]
        \hat{B}_-^{(\pm)} &= \mp\frac{i}{2}\sqrt{AI_-\frac{\mu_0\hbar\omega}{V_\mathrm{Cav}}}\qty[\hat{a}^\dagger e^{\mp i\qty(\frac{\omega}{c}z+\theta_-)} - \hat{a}e^{\pm i\qty(\frac{\omega}{c}z+\theta_-)}]. \label{SupplmentaryInfo:Eq:MagneticPlaneWave}
    \end{align}

    We derive Eq.~\ref{SupplmentaryInfo:Eq:LeftMirror_LossOperator} by using Eq.~\ref{SupplmentaryInfo:Eq:ElectricPlaneWave}~and~\ref{SupplmentaryInfo:Eq:MagneticPlaneWave} to compute the backward propagating Poynting vector. We then integrate it over the surface of the mirror to obtain the total impinging power and multiply the result by the mirror's transmission coefficient $\mathcal{T}$ to obtain the total transmitted power as the loss operator for the left end-mirror.

    \begin{equation}\label{SupplmentaryInfo:Eq:LeftMirror_LossOperator}
        \begin{split}
            \hat{W}_\mathcal{T}^\qty(\mathrm{Left}) = -\mathcal{T}\cdot & AI_-\cdot\frac{\hbar\omega c}{4L} \\[0.5\lineheight]
                                                                 \times & \qty[\hat{a}^\dagger e^{-i\qty(\frac{\omega}{c}\frac{L}{2}-\theta_-)} - \hat{a}e^{ i\qty(\frac{\omega}{c}\frac{L}{2}-\theta_-)}]^2
        \end{split}
    \end{equation}

    Similarly, the loss operator for the right end-mirror is found to be
    \begin{equation}\label{SupplmentaryInfo:Eq:RightMirror_LossOperator}
        \begin{split}
            \hat{W}_\mathcal{T}^\qty(\mathrm{Right}) = -\mathcal{T}\cdot & AI_+\cdot\frac{\hbar\omega c}{4L} \\[0.5\lineheight]
                                                                 \times & \qty[\hat{a}^\dagger e^{-i\qty(\frac{\omega}{c}\frac{L}{2}+\theta_+)} - \hat{a}e^{ i\qty(\frac{\omega}{c}\frac{L}{2}+\theta_+)}]^2.
        \end{split}
    \end{equation}

    The total contribution to the photon decay rate from the leakage through the end-mirrors is
    \begin{equation}\label{SupplmentaryInfo:Eq:EndMirror_DecayRate}
        \begin{split}
           \kappa_\mathcal{T}   &= \frac{1}{\hbar\omega(n+\frac{1}{2})}\ev**{\hat{W}_\mathcal{T}^\qty(\mathrm{Left})+\hat{W}_\mathcal{T}^\qty(\mathrm{Right})}{n} \\[0.5\lineheight]
                                &= \mathcal{T}\cdot\frac{c}{2L}\cdot A\qty(I_-+I_+).
        \end{split}
    \end{equation}

\subsection{Loss from Absorptive Membranes}
\label{SupplmentaryInfo:SubSec:AbsorptiveMembranes}

    To find the power dissipated within a membrane of complex refractive index $n+i\tilde{n}$, we integrate the Ohmic heating resulting from the membrane's conductivity. The imaginary component of a membrane's refractive index is related to it's conductivity through $\sigma = 2n\tilde{n}\epsilon_0\omega$~\cite{pedrottiIntroductionOptics2017}. The operator associated with the Ohmic heating caused by such a conductivity is readily calculated in Eq.~\ref{SupplmentaryInfo:Eq:AbsorptionLossOperator_Single}, with $\varphi\equiv \frac{nd\omega}{c}$.

    \begin{equation}\label{SupplmentaryInfo:Eq:AbsorptionLossOperator_Single}
        \begin{split}
           \hat{W}_\sigma   &=  \sigma \int_{V}\dd[3]{r} \hat{E}_0^2 \\[0.5\lineheight]
                            &=  -AI_0\cdot\frac{\hbar c\sigma}{n^3\epsilon_0 L}\cdot\qty(\hat{a}^\dagger-\hat{a})^2\cdot\xi(\varphi, \theta_0)
        \end{split}
    \end{equation}

    \begin{equation}\label{SupplmentaryInfo:Eq:AbsorptionGeometricFactor}
        \begin{split}
            \xi(\varphi, \theta)    &\equiv     \int_{0}^{\varphi}\dd{\tilde{z}}\cos[2](\tilde{z}+\theta) \\[0.5\lineheight]
                                    &=          \frac{1}{2}\qty(\varphi+\sin(\varphi)\cos(2\theta_0+\varphi))
        \end{split}
    \end{equation}

    For a system of multiple membranes, each membrane contributes a term similar to the total absorption loss operator in Eq.~\ref{SupplmentaryInfo:Eq:AbsorptionLossOperator_Single}. The overall photon decay rate due to absorption for a system of identical membranes\footnote{The case of non-identical membranes is handled by including membrane indices on $n$, $\tilde{n}$, and $\varphi$ to account for possible differences between membrane material and thickness.} in a general configuration is given by Eq.~\ref{SupplmentaryInfo:Eq:TotalAbsorptionLossOperator}. The total decay rate of the cavity is the sum of Eq.~\ref{SupplmentaryInfo:Eq:EndMirror_DecayRate}~and~\ref{SupplmentaryInfo:Eq:TotalAbsorptionLossOperator}.

    \begin{equation}\label{SupplmentaryInfo:Eq:TotalAbsorptionLossOperator}
        \kappa_\sigma = 4\frac{\tilde{n}c}{n^2 L}\cdot\sum_{i=1}^{N}AI_0^{(i)}\xi(\varphi, \theta_0^{(i)})
    \end{equation}

\subsection{Collective Coupling Efficiency Vs. Individual Coupling
Efficiency}
\label{SupplmentaryInfo:SubSec:CollectiveEfficiencyBound}

    We show now that a membranes individual coupling efficiency of Eq.~\ref{DecayRate:Eq:CouplingEfficiency_Max} also holds as an upper bound on the ratio of the collective coupling strength and the total photon decay rate of a system of $N$ identical membranes. We have that for any configuration, we may write the system's collective coupling strength $g_c$ and photon decay rate $\kappa$ as
    \begin{equation}\label{SupplmentaryInfo:Eq:CollectiveEfficiencyProof_Step1}
        \frac{g_c}{\kappa} = \frac{\sqrt{\qty(g^{(1)})^2+\qty(g^{(2)})^2+\dots+\qty(g^{(N)})^2}}{\kappa_\mathcal{T}+\qty(\kappa_\sigma^{(1)}+\kappa_\sigma^{(2)}+\dots+\kappa_\sigma^{(N)})}.
    \end{equation}
    For each membrane, we have $\abs*{g^{(i)}}\leq \eta_\mathrm{max}\cdot\kappa^{(i)}$, so
    \begin{equation}\label{SupplmentaryInfo:Eq:CollectiveEfficiencyProof_Step2}
      \frac{g_c}{\kappa} \leq  \eta_\mathrm{max}\cdot\frac{\sqrt{\qty(\kappa_\sigma^{(1)})^2+\qty(\kappa_\sigma^{(2)})^2+\dots+\qty(\kappa_\sigma^{(N)})^2}}{\kappa_\sigma^{(1)}+\kappa_\sigma^{(2)}+\dots+\kappa_\sigma^{(N)}}.
    \end{equation}
    Finally, for any set of positive numbers, the square of their sum is greater than or equal to the sum of their squares, yielding the desired inequality
    \begin{equation}\label{SupplmentaryInfo:Eq:CollectiveEfficiencyProof_Step3}
      \frac{g_c}{\kappa} \leq  \eta_\mathrm{max}.
    \end{equation}

\subsection{Beyond Simple Membranes}
\label{SupplmentaryInfo:SubSec:BeyondSimpleMembranes}

    The reflectivities of simple \enquote{slab} membranes are limited by the refractive index of the material. Designing membranes with photonic crystal structures~\cite{buiHighreflectivityHighQMicromechanical2012} or sub-wavelength gratings~\cite{stambaughMembraneinthemiddleMirrorinthemiddleHighreflectivity2015} can result in reflectivities very close to unity. The additions of such structures complicate the field profile within the membranes, making a first-principles calculation of the absorption, as was done for the simple membrane in Sec.~\ref{DecayRate:SubSec:AbsorptiveMembranes}, analytically intractable.

    As an effective treatment, we may consider the membrane to be an infinitely thin scatterer with complex polarizability $\xi+i\tilde{\xi}$~\cite{xuerebCollectivelyEnhancedOptomechanical2013, buiHighreflectivityHighQMicromechanical2012, spencerTheoryTwoCoupled1972}. The membrane's transfer matrix, reflection and transmission amplitude, and absorption coefficient are shown in Eq.~\ref{SupplmentaryInfo:Eq:NonSimpleMem_TransferMatrix}, \ref{SupplmentaryInfo:Eq:NonSimpleMem_ReflectTransmisCoefficient}, and \ref{SupplmentaryInfo:Eq:NonSimpleMem_AbsorpCoef} respectively.

    \begin{equation}\label{SupplmentaryInfo:Eq:NonSimpleMem_TransferMatrix}
        \vb{T} = \begin{pmatrix}
                    (1-\tilde{\xi})+i\xi    &   -\tilde{\xi}+i\xi   \\
                    \tilde{\xi}-i\xi        &   (1+\tilde{\xi})-i\xi
                 \end{pmatrix}
    \end{equation}

    \begin{align}
      r = \abs{\frac{-\tilde{\xi}+i\xi}{(1+\tilde{\xi})-i\xi}}  &&   t = \abs{\frac{1}{(1+\tilde{\xi})-i\xi}} \label{SupplmentaryInfo:Eq:NonSimpleMem_ReflectTransmisCoefficient}
    \end{align}

    \begin{equation}\label{SupplmentaryInfo:Eq:NonSimpleMem_AbsorpCoef}
        \mathcal{A} \equiv 1-r^2-t^2 = \frac{2\tilde{\xi}}{\qty(1+\tilde{\xi})^2+\xi^2}
    \end{equation}

    We work in the model of Fig.~\ref{SingleMembrane:Fig:SingleMembrane_Schematic} with $d=0$ and in the limit of weak losses $\qty(\frac{\tilde{\xi}}{\xi}\ll 1)$, and proceed to use the lossless mode profile\footnote{This amounts to setting $\tilde{\xi}=0$ in the membrane's transfer matrix $\vb{T}$.} in calculations of the coupling and decay rate. Using $\vb{T}$ to determine the membrane's intensity ratio, we find the coupling to be
    \begin{equation}\label{SupplmentaryInfo:Eq:NonSimpleMem_Coupling}
        g_0 = q_\mathrm{zpf}\cdot AI_-\frac{\omega}{L}\cdot 4\xi\qty[\xi\cos[2](\varphi_-)+\sin(\varphi_-)\cos(\varphi_-)],
    \end{equation}
    where $\varphi_-\equiv \frac{\omega}{c}(z-\frac{L}{2})+\theta_-$ is the phase of the left standing wave at the membrane.

    To determine the decay rate from absorption, we treat the imaginary component of the polarizability $\tilde{\xi}$ as originating from some surface conducting $\rho$ on the membrane. We recover $\vb{T}$ from the resulting wave equation only if
    \begin{equation}\label{SupplmentaryInfo:Eq:NonSimpleMem:SurfaceConductivity}
        \rho = \frac{2\tilde{\xi}}{\mu_0 c}.
    \end{equation}
    It is straightforward to find the energy dissipated from the Ohmic heating caused by this $\rho$. The resulting decay rate is
    \begin{equation}\label{SupplmentaryInfo:Eq:NonSimpleMem_AbsorpDecayRate}
        \kappa_\sigma = AI_-\cdot\frac{c}{L}\cdot4\tilde{\xi}\cos[2](\varphi_-).
    \end{equation}

    We note that both the coupling~(Eq.~\ref{SupplmentaryInfo:Eq:NonSimpleMem_Coupling}) and decay rate~(Eq.~\ref{SupplmentaryInfo:Eq:NonSimpleMem_AbsorpDecayRate}) vanish when the membrane is placed at a node of the cavity, as there is no field for the membrane to interact with. The vanishing of the decay rate is unphysical and arises only because we have taken the approximation of an infinitely thin membrane. In reality, the membrane has a finite thickness and hence will always have a non-vanishing interior field causing some energy loss.

    The coupling efficiency is given by
    \begin{equation}\label{SupplmentaryInfo:Eq:NonSimpleMem_Efficiency}
        \eta = 2\pi\frac{q_\mathrm{zpf}}{\lambda}\cdot\frac{\tilde{\xi}}{\xi}\cdot\abs{\xi+\tan(\varphi_-)}.
    \end{equation}
    This is singular when the decay rate vanishes, which as stated above is unphysical, as well as undesirable as the coupling also vanishes. In a configuration with appreciable coupling, $\tan(\varphi_-)$ will be $\mathcal{O}(1)$. Neglecting this singular term, our condition for the possibility of strong single-photon coupling is
    \begin{equation}\label{SupplmentaryInfo:Eq:NonSimpleMem_StrongCouplingCondition}
        2\pi\frac{q_\mathrm{zpf}}{\lambda}\cdot\frac{\xi^2}{\tilde{\xi}} > 1.
    \end{equation}

    It is convenient to recast Eq.~\ref{SupplmentaryInfo:Eq:NonSimpleMem_StrongCouplingCondition} in terms of the membrane's reflectivity $\mathcal{R}\equiv r^2$ and absorption coefficient $\mathcal{A}$. The expressions for $\mathcal{R}$ and $\mathcal{A}$ to lowest non-vanishing order in $\frac{\tilde{\xi}}{\xi}$ are given in Eq.~\ref{SupplmentaryInfo:Eq:NonSimpleMem_ReflecAbsorp}. The strong coupling condition is shown in Eq.~\ref{SupplmentaryInfo:Eq:NonSimpleMem_StrongCouplingCondition_ReflectAbsorp}.

    \begin{align}
        \mathcal{R} = \frac{\xi^2}{1+\xi^2}     &&   \mathcal{A} = \frac{2\tilde{\xi}}{1+\xi^2} \label{SupplmentaryInfo:Eq:NonSimpleMem_ReflecAbsorp}
    \end{align}

    \begin{equation}\label{SupplmentaryInfo:Eq:NonSimpleMem_StrongCouplingCondition_ReflectAbsorp}
        \frac{\mathcal{A}}{\mathcal{R}} < 4\pi \frac{q_\mathrm{zpf}}{\lambda}
    \end{equation}

    Eq.~\ref{SupplmentaryInfo:Eq:NonSimpleMem_StrongCouplingCondition_ReflectAbsorp} is a more general condition for the possibility of strong single-photon coupling than what was found in Sec.~\ref{DecayRate:SubSec:LimitationsAbsorptiveMembranes} and is valid so long as the description of the membrane as an infinitely thin scattering element is acceptable.

\newpage
\bibliography{bibliography}
\end{document}